\documentclass[twocolumn,letterpaper,amsmath,amssymb,amsfonts,floatfix,aps,superscriptaddress,showkeys,showpacs,nofootinbib]{revtex4-1}
\usepackage{bm}
\usepackage{ifpdf}
\usepackage{epsfig}
\usepackage{color}
\usepackage{multirow}
\usepackage{float}
\usepackage{graphicx}
\usepackage{amsmath,amssymb}
\usepackage[makeroom]{cancel}

\newcommand{\kbt}{k_{\mathrm{B}}T}

\begin{document}

\title{Charge Regulation with Fixed and Mobile Charges}
\author{Yael Avni}
\affiliation{Raymond and Beverly Sackler School of Physics and Astronomy\\ Tel Aviv
University, Ramat Aviv, Tel Aviv 69978, Israel}
\affiliation{School of Physical Sciences, University of Chinese Academy of Sciences, Beijing
100049, China}
\author{David Andelman}
\email{andelman@post.tau.ac.il}
\affiliation{Raymond and Beverly Sackler School of Physics and Astronomy\\ Tel Aviv
University, Ramat Aviv, Tel Aviv 69978, Israel}
\author{Rudolf Podgornik}
\email{podgornikrudolf@ucas.ac.cn. Also affiliated to: Department of Theoretical Physics, Jo\v zef Stefan Institute, SI-1000 Ljubljana, Slovenia and Department of Physics, Faculty of Mathematics and Physics, University of Ljubljana, SI-1000 Ljubljana, Slovenia.}
\affiliation{School of Physical Sciences, University of Chinese Academy of Sciences, Beijing
100049, China}
\affiliation{Kavli Institute for Theoretical Sciences, University of Chinese Academy of Sciences, Beijing 100049, China, and CAS Key Laboratory of Soft Matter Physics, Institute of Physics, Chinese Academy of Sciences, Beijing 100190, China}

\begin{abstract}
Uncompensated charges do not occur in Nature and any local charge should be a result of charge separation. Dissociable chemical groups at interfaces in contact with ions in solution, whose chemical equilibrium depends both on short-range non-electrostatic and long-range electrostatic interactions, are the physical basis of this charge separation, known as {\em charge regulation} phenomena. The charged groups can be either fixed and immobile, as in the case of solvent-exposed solid substrate and soft
bounding surfaces, ({\em e.g.}, molecularly smooth mica surfaces and soft phospholipid membranes), or free and mobile, as in the case of charged macro-ions, ({\em e.g.}, protein or other biomolecules). Here, we review the mean-field formalism used to describe both cases, with a focus on recent advances in the modeling of mobile charge-regulated macro-ions in an ionic solution. The general form of the screening length in such a solution is derived, and is shown to combine the concept of intrinsic capacitance, introduced by Lund and J\"{o}nsson, and bulk capacitance, resulting from the mobility of small ions and macro-ions. The advantages and disadvantages of different formulations, such as the cell model vs. the collective approach, are discussed, along with several suggestions for future experiments and modeling.

\end{abstract}
\maketitle

\section{Introduction}

The mechanism of ion exchange between dissociable amino acids and their surrounding solution has been proposed already in the 1920's by Linderstr{\o}m-Lang of the Carlsberg Laboratory~\cite{Linderstrom-Lang}. Early advances in the dissociation/association equilibria and acid-base properties of polyelectrolytes~\cite{Lifson, Marcus} were introduced in the groundbreaking works of Kirkwood and Schumaker~\cite{KirkSchu}.
These, as well as the dissociation equilibria of proteins (Tanford and Kirkwood~\cite{Tanford}), have been authoritatively reviewed by Borkovec, J\"{o}nsson, and Koper~\cite{borkovec2001ionization}.

Another important contribution came in the 1970's when Ninham and Parsegian~\cite{NP-regulation} introduced the {\em charge regulation} (CR) mechanism. In their  seminal work, they developed a self-consistent relationship between the local electrostatic potential and the dissociated state of chargeable surface groups. The novelty at that time was to introduce a more special charge-regulated boundary conditions for the Poisson-Boltzmann (PB).
The charge association/dissociation process (CR mechanism) couples the local electrostatic field with the local charge, and results in a self-consistent partitioning of dissociated and associated surface states~\cite{Safinya,Markovich2016EPL}.

Charge regulation governs many electrostatic interactions in biological systems, making it fundamental to the understanding of protein complexation~\cite{ProteinRNA2018, ProteinNanoparticleBinding} and adsorption onto surfaces~\cite{Hartvig2011, Hyltegren2017}, bacterial adhesion~\cite{CRbacterialcell2008}, viral capsids assembly~\cite{capsid2014}, translocation of DNA through solid-state nanopores~\cite{nanopore2011} and several other bio-processes~\cite{Perutz1978}. It is also a key ingredient for the design of materials based on polyelectrolytes in solution and polyelectrolyte brushes~\cite{Netz2003, Borukhov2000, Kumar2012}.

The CR formulation can be implemented via the law of mass action~\cite{Regulation4, Borkovec2015}, and separately, by modifying the surface part of the total free-energy~\cite{CTAB,  Pincus, Olvera, Olvera2, Natasa1, Maggs, Diamant, Ben-Yaakov2}. The latter approach leads to the same results as the law of mass action, but with the advantage that it can be easily generalized to include any non-electrostatic surface interactions~\cite{Harries2006, Majee2018}.

The Poisson-Boltzmann theory with CR surfaces has been studied in the past for uniform charge distributions of dissociable groups, in contact with an electrolyte solution~\cite{chan1976electrical, Borkovec2008, Markovich2016EPL}. Other studies involved modeling of a single CR colloid in solution, in the proximity of another charged surface~\cite{PhysRevLett.117.098002}. Most of previous calculations employed linearized CR boundary conditions or a linearized version of the PB equation itself (known as the Debye-H\"uckel limit)~\cite{Borkovec1, Borkovec2, Carnie2, Healy1978}. The assumption of uniform charge distribution was dropped in later works~\cite{Borkovecpatchy2010, Boonpatchy}, and CR surfaces with {\em patchy chargeable groups} were analyzed as well.
In some cases, it was found that higher-order electrostatic multipoles may need to be considered in relation to the CR process, in addition to the monopolar ones~\cite{Anze1, Lund2016}.

A number of models of single protein and protein-protein interactions~\cite{Leckband2001} in aqueous solution have been studied by various simulation techniques~\cite{Lund2005,Lundproteinads, Lund2005,Fer2009, Warshel2006, Teixeira2010}, and extensively reviewed in Ref.~\cite{Lundreview}. For proteins, the CR contributes significantly to the fluctuation part of the electrostatic interaction (the Kirkwood-Schumaker interaction~\cite{KirkSchu}). This contribution can be quantified in terms of the charge capacitance, which is a measure of the molecular charge fluctuations defined by the variance of the mean charge~\cite{Lundchamaleon}.

In order to evaluate the importance of charge regulation in protein and polyelectrolyte systems, the standard MC algorithm has to be augmented in order to account for the protonation/deprotonation reaction of the acidic/basic sites. This implies an additional MC step with energy change that follows the Nerst equation, $\Delta U  = \Delta U_{\rm ES} \pm \kbt \log_{10}\mathrm{(pH - pK_0)}$, where $\Delta U_{\rm ES}$ is the change in the standard electrostatic energy~\cite{ullner1994conformational}.

Charge regulation was shown to have pronounced effects on the properties of weak polyelectrolytes, such as the pH-dependence of chain conformation and ionization~\cite{Borukhov2000, Olvera, Olvera2}. This has been explored for
linear~\cite{ullner1994conformational}, star-like~\cite{Uhlik2014} and macroscopic networks~\cite{Rud2017,Olvera3}, and microgel architectures~\cite{Schneider2018} of weak polyelectrolytes, and for physically deposited and chemically grafted polyelectrolyte layers~\cite{Olvera,Olvera4}.

The PB theory of  a solution of coupled CR macro-ions was addressed to a much lesser extent, mostly within the context of the {\em cell model}~\cite{Alexander1984}, for
which each macro-ion is placed in the center of a cell whose external boundary mimics the presence of neighboring macro-ions. In this way, the interactions between the macro-ions are taken into account on a simple mean-field level~\cite{borkovec2001ionization}. The cell model was later generalized to include charge regulation of macro-ion surfaces~\cite{Gisler1994}, which allowed to find the macro-ion effective charge as a function of their concentration~\cite{boon2012charge}, and the phase behavior of oppositely charged macro-ion mixtures~\cite{Allen2004, Biesheuvel2004Langmuir, Biesheuvel2006}.

\begin{figure*}
\includegraphics[width = 2.05\columnwidth,draft=false]{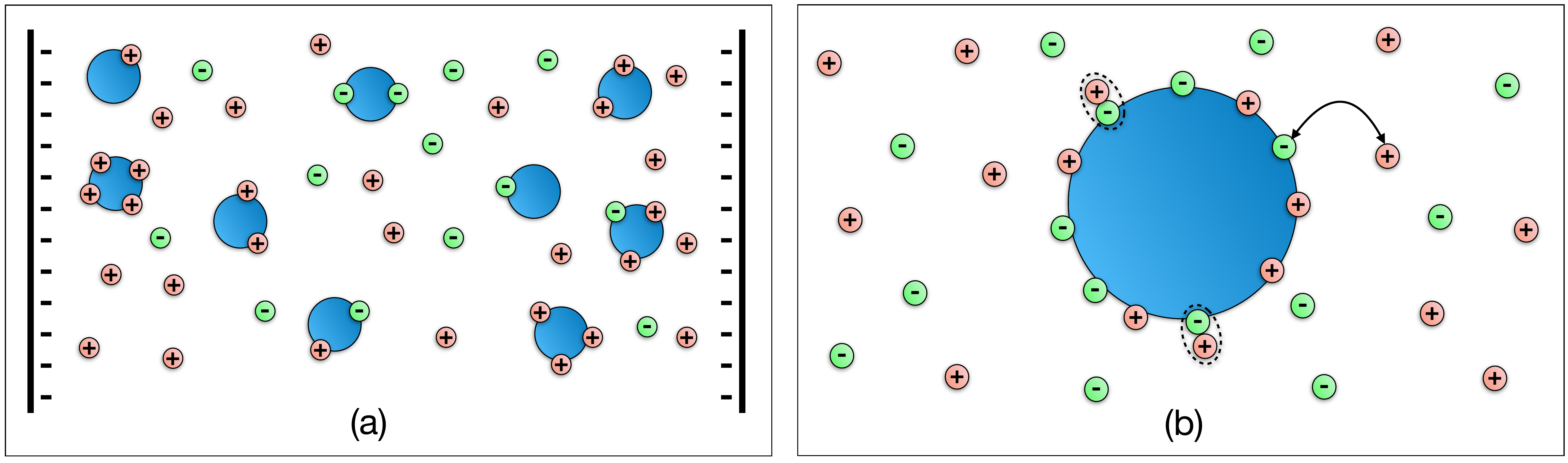}
\caption{\textsf{(color online) Schematic drawing of (a) a solution containing simple salt ions (red and green) and charge regulated macro-ions (blue), in the vicinity of two negatively charged surfaces; (b) the charge regulation process on a single macro-ion, for which the positive salt ions are free to adsorb/desorb onto the macro-ions. The extent of adsorption/desorption is determined by the local electrostatic potential, causing both the macro-ion effective charge and macro ion concentration to vary in space.}}
\label{Fig1}
\end{figure*}

While the cell model gives a reasonable approximation for the effective charge of the macro-ions in the homogeneous bulk, it cannot describe the collective effects due to external electric fields, where both the effective charge and macro-ion concentration vary in space, as is depicted in Fig.~\ref{Fig1}. To account for such effects, one needs to employ a more refined and collective description.

An attempt in that direction was done in a study of the sedimentation of CR colloids~\cite{Biesheuvel2004}. However, as the theory was not derive from first principles, its consistency remains uncertain. To that end, a general formalism was introduced in Refs.~\cite{Tomer2017,Yael2018}, and  account for mobile macro-ion effects in dilute solutions. The macro-ions are treated as point-like particles, similar to small salt ions, while retaining their internal degrees of freedom that determine the macro-ions charge state in a self-consistent way.

In this short review, we discuss some recent developments in the theoretical modeling of CR macro-ions in solution. First, the general free-energy formalism is introduced in section II. Then, we review the cell model (section III) and concentrate on the collective description of mobile macro-ions (section IV), as was set forward in Refs.~\cite{Tomer2017,Yael2018}. We end the review (section V) by offering several concluding remarks and future prospects. Throughout this review, we shall employ the mean-field formalism and neglect any fluctuation effects, which are addressed elsewhere~\cite{Lundreview,Natasa1, Natasa2}.

\section{The General Free-Energy Formalism for Charge Regulation}\label{sec:approach}
The PB mean-field theory of charge-regulating processes can be formulated as a density functional theory of the free energy. The free-energy functional consists of bulk and surface terms, and
the combined free energy assumes the form
\begin{eqnarray}
\label{density function}
{\cal  F} &=& \int_{_{\rm V}}\text{d}^{3}r\, f_{_{\rm V}}(\psi, \nabla\psi, n_i) + \int_{\rm s}\text{d}^{2}r \, f_{\rm s}(\psi_{\rm s}, n^{\rm s}_i)\, ,
\end{eqnarray}
where $\psi({\bf r})$ is the local electrostatic potential, $n_i({\bf r})$ is the local concentration of species $i=1,2,\dots$, and their values on the surface are denoted as $\psi_{\rm s}$ and $n^{\rm s}_i$, respectively. The volume free-energy, $f_{_{\rm V}}$, contains the electrostatic energy and free-energy terms of an uncharged system.  In a dilute solution, these terms represent the ideal-gas entropy of the ions. The surface part,
$f_{\rm s}$, essentially includes the CR effect via the ion-surface interaction, {\em i.e.}, charge regulation.

Assuming that the system is composed of monovalent salt, then
\begin{equation} \label{fPB}
f_{_{\rm V}}(\psi, \nabla\psi, n_i) = f_{_{\rm PB}}(\psi, \nabla\psi, n_{\pm}) \, ,
\end{equation}
is the standard PB free-energy density~\cite{Safinya}. For simplicity, we further assume  that the only ion type that exchanges at the surface is one of the salt species, chosen to be the cation. An additional assumption is that only a single adsorption/desorption mechanism is involved.
Consequently, $f_{\rm s}$, is a function of the surface potential $\psi_{\rm s}$, and of the surface concentration of the adsorbed cations, $n^{\rm s}_{+}$. Its exact functional form depends on the CR model that is employed~\cite{Harries2006}.

The volume part of the Euler-Lagrange equations is given by
\begin{equation}
\label{formalism1}
\nabla \cdot \frac{\partial f_{_{\rm V}}}{\partial \nabla \psi} - ~\frac{\partial f_{_{\rm V}}}{\partial  \psi} = 0
~~~{\rm and}~~~
\frac{\partial f_{_{\rm V}}}{\partial n_{\pm}} = 0 \, ,
\end{equation}
while the surface part is
\begin{equation}
\label{formalism2}
\left. {\bf n}\cdot\frac{\partial f_{_{\rm V}}}
{\partial \nabla \psi} \right|_{\rm s}  + ~\frac{\partial f_{\rm s}}{\partial \psi_{\rm s}} = 0
~~~{\rm and}~~~
 \frac{\partial f_{\rm s}}{\partial n^{\rm s}_+} = 0,
\end{equation}
where $\bf n$ is a unit vector normal to the bounding surface(s). Equation~(\ref{formalism1}) reduces to the PB equation, and Eq.~(\ref{formalism2}) yields the exact CR boundary condition, which was originally derived using chemical equilibrium equation~\cite{NP-regulation}. The above formalism can be extended to describe a variety of other geometries, with the only limitation being that the boundaries are taken to be fixed (immobile) in space.

\section{The Cell-model for Charge-Regulated macro-ions}
As mentioned in the introduction, a viable way to describe charge regulation of immobile macro-ions
is the cell-model approach. Each macro-ion occupies the center of an imaginary Wigner-Seitz cell, and is surrounded by solvent molecules and salt ions. Both the cell and the central macro-ion are taken to have a spherical shape for simplicity (although a cylindrical cell is used to model polyelectrolytes in solution).

The macro-ion fixed radius is denoted by $a$, while the cell radius, $R$, is determined by the concentration (per unit volume) of macro-ions, $p$, such that $R \sim p^{-{1}/{3}}$. We can now apply the formalism presented in section~\ref{sec:approach}, with the additional demand of electro-neutrality in each cell, separately. For the bulk part, Eq.~(\ref{formalism1}) remain the same and the CR boundary is described by Eq.~(\ref{formalism2}), while at the outer cell boundary, the additional boundary condition is
\begin{equation}
{\bf n}\cdot {\bf E}\Big|_R =   {\bf n}\cdot \frac{\partial f_{_{\rm V}}}{\partial \nabla \psi}\Big|_R = 0,
\end{equation}
as is stipulated by symmetry.

Solving these equations, one can derive ionic profiles, effective macro-ion charges and electrostatic pressure as function of the macro-ion concentration $p$,
while the interactions between the macro-ions are taken into account in an indirect manner, via the external boundary condition at $R$.

This single-particle cell-model approach is mostly appropriate at high density of the macro-ions, where their translational entropy is small or even vanishing~\cite{Alexander1984}, or when one wants to describe a homogeneous bulk. However, it cannot describe collective effects such as the response of  macro-ions in solution to external fields.

\section{The Collective approach for Charge Regulated macro-ions}
In the collective approach, applicable in the limit of dilute macro-ion solutions, the macro-ions themselves are treated in analogy to point-like salt ions~\cite{Tomer2017, Yael2018}. Assuming a solution containing many point-like CR macro-ions with concentration $p$, the surface term in  Eq.~(\ref{density function}) vanishes and is replaced by additional terms in the volume part of the free energy. Equation~(\ref{fPB}) now reads,
\begin{equation}
f_{_{\rm V}}(\psi, \nabla\psi, n_\pm, p, Q_p) =  f_{\rm PB}(\psi, \nabla\psi, n_{\pm}, p) + p\,g(\psi, Q_p).
\label{fcelw}
\end{equation}
where $Q_p$ is the overall macro-ion charge, and $g(\psi, Q_p)$ is the point-like version of the former $f_{\rm s}(\psi_{\rm s}, n_{+}^{\rm s})$, satisfying
\begin{eqnarray}
g({\bf r}) &= & \lim_{a\rightarrow 0}  \int_{\rm s}\text{d}^{2}r\,
f_{\rm s}(\psi_{\rm s}({\bf r}), n^{\rm s}_{+}({\bf r}))\nonumber\\
& =& g(\psi({\bf r}),Q_p({\bf r})) \, .
\end{eqnarray}
The volume part of Eq.~(\ref{fcelw}) is now composed of the PB free energy of the three ionic species, $n_\pm$ and $p$, and the bulk CR term $p\, g(\psi, Q_p)$.

In this formalism, Eq.~(\ref{formalism1}) remain the same, but they are complemented not by Eq.~(\ref{formalism2}), but by two other conditions,
\begin{equation}
\label{formalism3}
\frac{\partial f_{_{\rm V}}}{\partial p} = 0 \quad~ {\rm and} \quad~ \frac{\partial f_{_{\rm V}}}{\partial Q_p} = 0.
\end{equation}
The macro-ions are now described on the same footing as the solution salt ions, except that their charge, $Q_p$, is not fixed, but is determined self-consistently.

The most important difference between the cell model and the collective description is the macro-ion translational entropy described by $f_{\rm PB}(\psi, \nabla\psi, n_{\pm}, p)$. In addition, the CR affects the charge of the macro-ion $Q_p$, as well as its concentration $p$ and the corresponding electrostatic potential. The latter is averaged over a local distribution of macro-ions and that of the salt ions around the macro-ions.

\begin{figure}
\includegraphics[width = 1\columnwidth,draft=false]{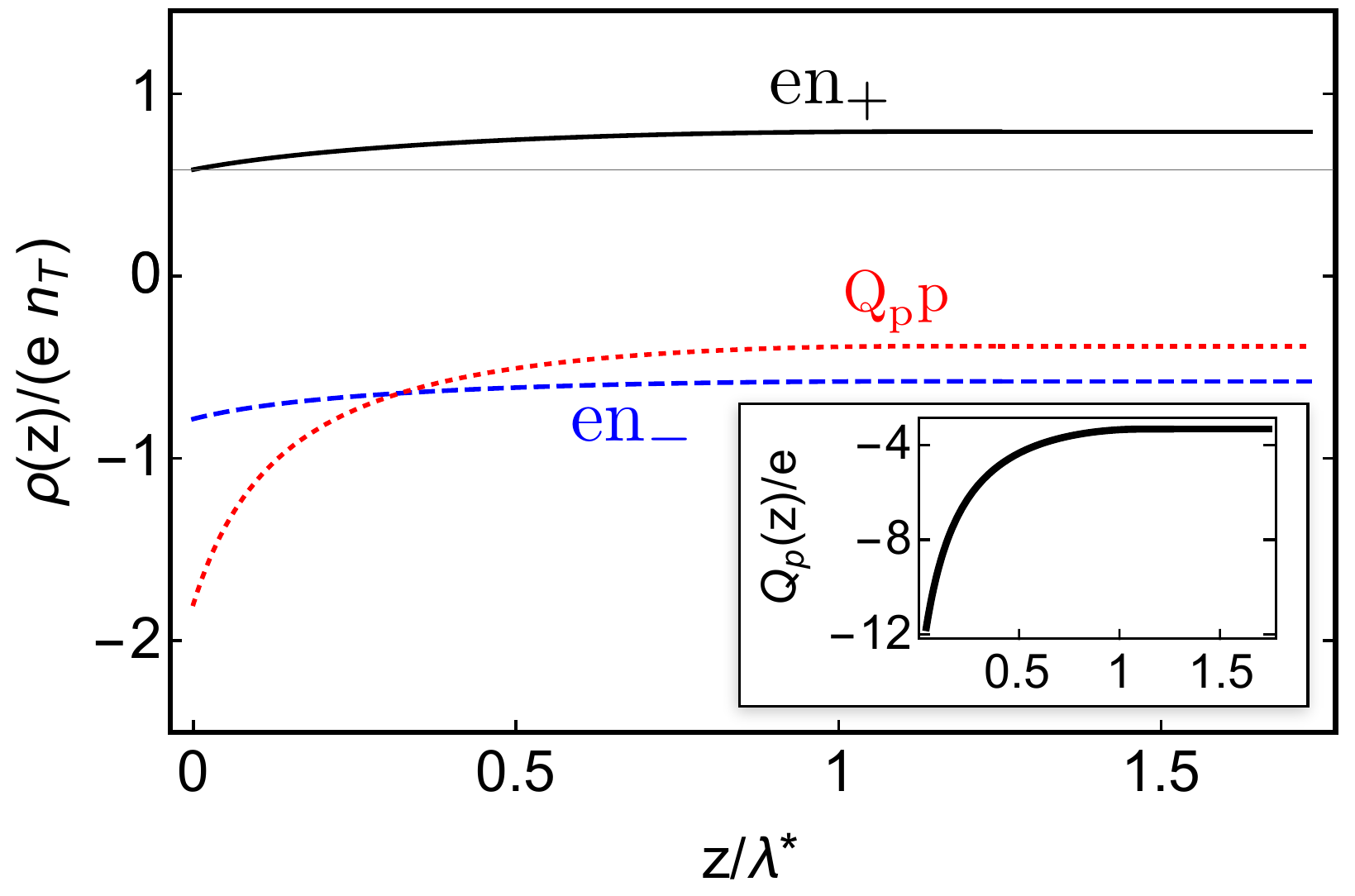}
\caption{\textsf{The three contributions to the total charge density, $\rho(z) = en_{+}-en_{-}+pQ_{p}$  as function of the distance from a positively charged surface: cation charge density $e n_+$ (solid black line), anion charge density ${-}e n_-$ (dashed blue line) and macro-ion charge density $pQ_p$ (dotted red line). $\rho(z)$ is normalized by the unit charge $e$ and the total salt concentration $n_{\rm T}$, and the distance is normalized by the length scale $\lambda^* = 1/\sqrt{8\pi l_{\rm B} n_{\rm T}}$.  Note that the total amount of ions, $n_{\rm T}$, includes the free and adsorbed ions and is different than $n_b$. The total concentration and its corresponding $\lambda^{*}$ are used here merely to rescale charge concentration and distances.
The macro-ion charge is regulated by a simple mechanism described in ref.~\cite{Tomer2017}. Inset: the macro-ion effective charge $Q_{p}(z)$ normalized by the unit charge $e$ as function of the distance, $z/\lambda^*$, from the surface. Results adapted from ref.~\cite{Tomer2017}
}}
\label{Fig2}
\end{figure}

The mobility of the macro-ions
has several important consequences. In the presence of external fields, both the macro-ion concentration ($p$) and charge ($Q_p$) vary in space, as is demonstrated in Fig.~\ref{Fig2}. Some insights on this behavior can be obtained by looking at the effective screening length of the system, $\lambda_{\rm eff}$,
\begin{equation}
\label{screening}
\lambda_{\rm eff}^{-2} = -\frac{1}{\varepsilon_0 \varepsilon} \frac{\partial\rho(\psi)}{\partial\psi} \Big|_{\psi=0} \,  ,
\end{equation}
defined in analogy with the Debye screening length, $\lambda_{\rm D}$,
where $\rho$ is the local charge density, to be defined below.

We recall that the charge density for an electrolyte solution having $N$ ionic species, each with bulk concentration $n_{\rm b}^i$ and with constant charge $q_i$ on each ionic species $i$ is:
$\rho(\psi)= \sum_{i}q_i n_i(\psi)$, where $n_{i}\left(\psi\right)=n_{\rm b}^{i}\exp\left(- q_{i}\psi/k_B T\right)
$, and the Debye screening length is given by $\lambda_{\rm D}^{-2} = \sum_{i}q_i^2 n_{\rm b}^i/(\varepsilon_0\varepsilon\kbt)$.
For a monovalent solution, the Debye length reduces to $\lambda_{\rm D} = 1/\sqrt{8\pi l_{\rm B} n_{b}}$ where $l_{\rm B}$ is the Bjerrum length $l_{\rm B} = e^2/(4\pi\varepsilon_0\varepsilon\kbt)$.

In the CR case, the screening length depends similarly on $\rho(\psi)$, the local charge density, but as the macro-ions are charge regulated, the expressions are somewhat different:
\begin{eqnarray}
\rho(\psi) &=& e n_{+}(\psi) -e n_{-}(\psi)+ Q_p(\psi)p(\psi).
\end{eqnarray}
and
\begin{equation}
\label{screening2}
\lambda_{\text{eff}}^{-2}=  \frac{4\pi l_{\rm B}}{e^2}\left[ e^2 n_{b}^{+}+ e^2 n_{b}^{-}+p_{b}\left({\overline{Q}_p}^{2}+ {(\triangle Q_p)^2}\right)\right],
\end{equation}
where $\overline{Q}_p \equiv Q_p(\psi {=} 0)$ is the average and the fluctuations around the average, due to charge regulation, are given by
\begin{equation} \label{meanvariance}
(\triangle Q_p)^2 \equiv - \kbt\frac{\partial Q_p(\psi)}{ \partial \psi}\Big|_{\psi=0}.
\end{equation}
The above relations for the average macro-ion charge and its variance remain the same for all CR models, while the specific form of $\overline{Q}_p$ and $\triangle Q_p$ may vary according to the CR model in mind.

An interesting feature of Eq.~(\ref{screening2}) is that it allows us to understand the system in terms of the overall capacitance, {\em i.e.}, the charge density response to the imposed variation in the electrostatic potential. The capacitance of the system has two contributions: the {\it bulk} capacitance, stemming from the spatial redistribution of the charged particles ($n_{\pm}$ and $p$), and the {\it intrinsic} capacitance due to the ability of each of the CR particles  to adjust its charge, $Q_p$. Hence, simple ions having a fixed charge, contribute only to the bulk capacitance, whereas the macro-ions contribute to both. Note that the intrinsic capacitance is the same as the capacitance defined by Lund and J\"{o}nsson~\cite{Lundreview} in order to quantify the Kirkwood-Schumaker interaction. Furthermore, Eq.~(\ref{screening2}) is a generalization of that capacitance for the mobile macro-ion case.

When using a theory that does not  take collective effects properly into account, one might derive Eq.~(\ref{screening2}) without the important $\triangle Q_p$ term~\cite{Bartlett}.
However, this term may modify the screening substantially, especially if the macro-ions are close to the point of zero charge, such that the $\overline{Q}_p$ term is small.

Different forms of $g(\psi)$ depend on the specific model under investigation, and give rise to a rich variation in the screening length, and consequently affect the electrostatic properties.
In particular, it is possible to classify CR mechanisms by their {\em asymmetry} between the positive and negative ions adsorption~\cite{Tomer2017}.
In this review, we assumed for simplicity that only cations can be adsorbed on the macro-ions, but the generalization to include a second adsorption process for the anions is straightforward as is presented in Refs.~\cite{Markovich2016EPL, Tomer2017, Yael2018}. In symmetric CR models, the macro-ions adsorb or release cations and anions in the same amount, resulting in a preference to an overall zero macro-ion charge, $\overline{Q}_p=0$. Asymmetric models, on the other hand, lead to highly charged macro-ions.

\begin{figure}
\includegraphics[width = 1\columnwidth,draft=false]{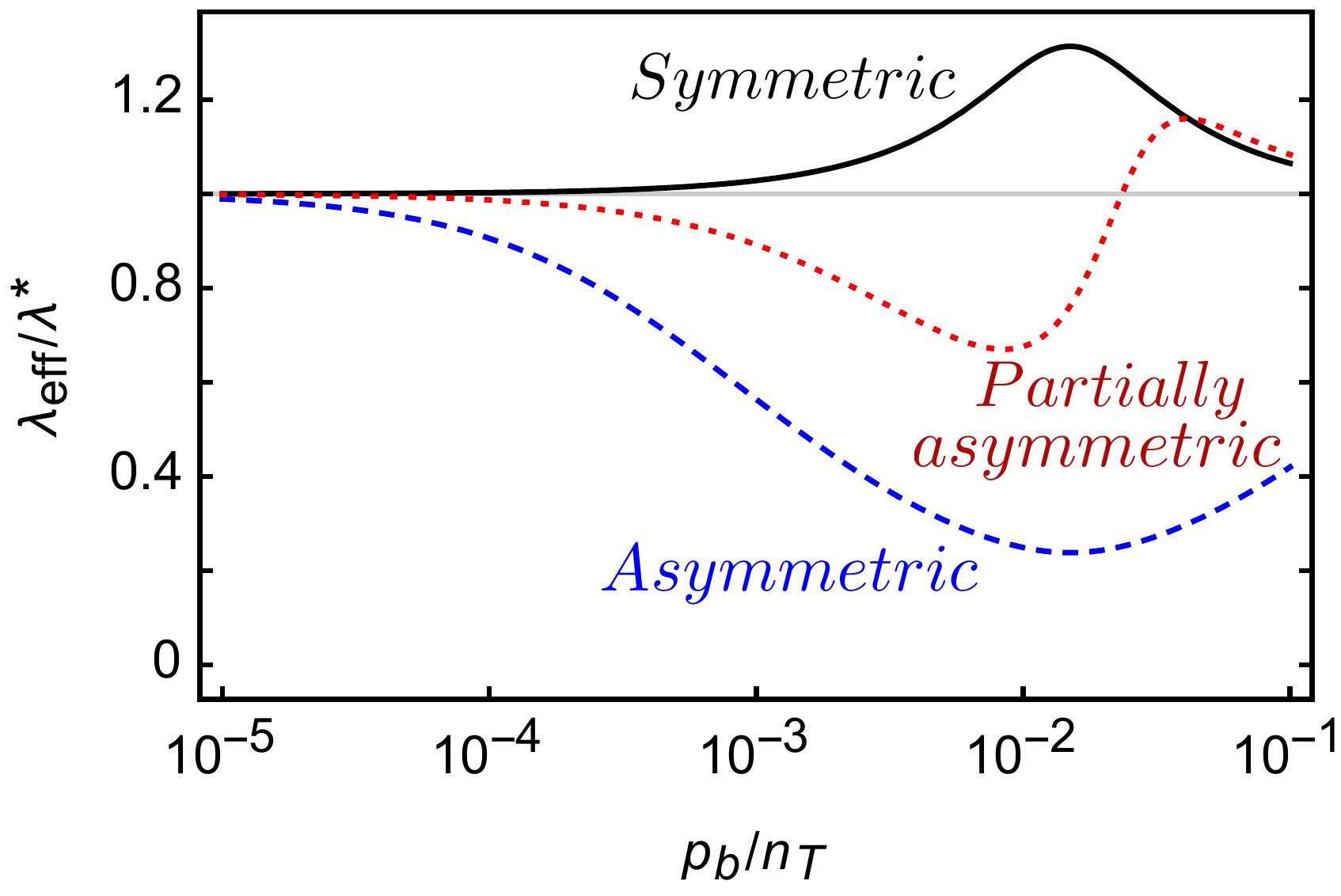}
\caption{\textsf{The effective screening length $\lambda_{\rm eff}$ (Eq.~(\ref{screening2})) normalized by the Debye length $\lambda^* = 1/\sqrt{8\pi l_{\rm B} n_{\rm T}}$, plotted as function of the macro-ion bulk concentration $p_b$ normalized by the total salt concentration $n_{\rm T}$.
The three curves represent three different CR mechanisms with a common theme: the macro-ions can adsorb both cations and anions from the solution via two distinct adsorption processes. The black solid curve represents the case where the two adsorption cancel one another such that the overall macro-ion charge is zero (symmetric case), whereas the dashed blue and dotted red curves represent cases where one adsorption is more dominant than the other, dashed blue being the more extreme case (asymmetric cases). Results adapted from Ref.~\cite{Yael2018}.
}}
\label{Fig3}
\end{figure}

The asymmetry is determined by the model parameters, for example the number of positive/negative dissociable groups, and the free energy gain from each dissociation. Symmetric and asymmetric models lead to very different dependence of the screening length on the macro-ion concentration, as was shown in Ref.~\cite{Yael2018} (see Fig.~\ref{Fig3}). Apart from the models discussed in Ref.~\cite{Yael2018}, other classification and more complicated CR mechanisms surely exist, their rich behavior awaiting to be uncovered.

\section{Concluding Remarks and Future prospects}

Our understanding of charge regulation phenomena grew immensely in the passing decades since its discovery~\cite{Linderstrom-Lang} and first rigorous formulation~\cite{NP-regulation}. Most of the attention has been given to the problem of a single CR macromolecule in solution, or the interaction between two CR macromolecules or between two CR surfaces~\cite{borkovec2001ionization,ullner1994conformational, borkovec1997difference,Lundproteinads, Borkovec1}.
However, the many-body problem of multiple and coupled CR macromolecules was addressed to a much lesser extent, being mainly analyzed within two frameworks: the cell model~\cite{Alexander1984, Gisler1994} and the collective approach~\cite{Tomer2017, Yael2018}, each having its own advantages and merits in different situations.

The collective description of mobile CR  macro-ions in an electrolyte solution is a simple generalization of the PB paradigm to the case of more complex ionic solutions, composed of macro-ions or nano-particles with non-trivial association/dissociation properties. It is particularly useful in order to understand the screening properties of the bulk solution as well as the inhomogeneous density and electrostatic potential distribution close to an externally imposed charge, as in the case of a bounding charged interface.

Such a collective description provides an explicit generalization of the screening length that consistently takes into account the redistribution of charge density of all the mobile charged species (just as is done for the standard Debye screening length), combined with specific changes in the macro-ion charge due to the charge regulation process itself. These two properties together determine the screening response of the solution. The understanding of the screening phenomenon in such complex ionic solutions should have repercussions not only conceptually but also practically when decay lengths are extracted from experimental data and compared to theoretical predictions.

The possibly high values of the macro-ion charge ($Q_p$), which can reach up to hundreds of unit charges, make the applicability range of the collective approach difficult to delimit. As the charge of the macro-ions itself varies, a systematic electrostatic coupling constant expansion that defines the strong coupling limit~\cite{Netz2001} is hard to obtain, and further detailed testing of the collective approach either by new experiments and/or detailed simulations is therefore highly desirable.

The collective approach may see other important applications in the future, particularly in the investigations of inhomogeneous systems subjected to external fields, such as centrifugal sedimentation of colloids~\cite{Biesheuvel2004}, proteins near charged membranes, or protein near impermeable membranes that generate electric fields due to the Donnan equilibrium.

On the theoretical side, it would be interesting to study different and perhaps more detailed CR models, including interactions between different dissociable groups, within the collective approach. Such short-range non-electrostatic interactions will generate higher-order terms in the free energy, and may lead to yet unexplored phase separations and phase transitions. Hopefully, these and other unsolved issues will be addressed in the future by experiments and theory.

{\em Acknowledgement:~~~} This research has been supported by the Naomi
Foundation through the Tel Aviv University GRTF Program. One of us (YA) is grateful for the hospitality of the University of the Chinese Academy of Sciences and the Institute of Physics, CAS, where part of this work has been conducted. RP would like to acknowledge the support of the {\em 1000-Talents Program} of the Chinese Foreign Experts Bureau, and DA acknowledges partial support from the ISF-NSFC (Israel-China) joint program under Grant No. 885/15.



\end{document}